\documentclass[twocolumn,prl,showpacs,floatfix]{revtex4}
\usepackage{graphicx}
\usepackage{epsfig}
\usepackage{rotating}
\usepackage{amsmath}
\usepackage{amsfonts}
\usepackage{amssymb}
\usepackage{enumerate}
\usepackage{longtable}
\usepackage{dcolumn}
\usepackage{bm}

\begin{document}

\title{Corrections to Pauling residual entropy and
single tetrahedron based approximations for the pyrochlore lattice Ising
antiferromagnet
}

\author{R. R. P. Singh}
\affiliation{Department of Physics, University of California Davis, CA 95616, USA}

\author{J. Oitmaa }
\affiliation{School of Physics, The University of New South Wales,
Sydney 2052, Australia}

\date{\rm\today}

\begin{abstract}
We study corrections to single tetrahedron based approximations
for the entropy, specific heat and uniform susceptibility of the
pyrochlore lattice Ising antiferromagnet, by a
Numerical Linked Cluster (NLC) expansion.
In a tetrahedron based NLC, the first order gives the Pauling residual
entropy of ${1\over 2}\log{3\over 2}\approx 0.20273$.
A $16$-th order NLC calculation changes the residual entropy 
to $0.205507$ a correction of $1.37$ percent over the Pauling value.
At high temperatures,
the accuracy of the calculations is verified 
by a high temperature series expansion. We find the
corrections to the single tetrahedron approximations to 
be at most a few percent
for all the thermodynamic properties.
\end{abstract}


\maketitle

\section{Introduction}

The residual entropy of ice was first calculated in a
classic paper by Pauling.\cite{pauling} This entropy arises
from the fact that an oxygen atom in ice is surrounded by
four protons. Any two of them
can move in close to it, making up the $H_2O$ unit, while the other
two stay away becoming part of the neighboring water molecules.
Extensive ground state
entropy is a hallmark of highly frustrated systems, and is well
appreciated in context of Ising models, at least since the exact solution of
the triangular-lattice antiferromagnet.\cite{wannier}
A direct connection between the entropy of ice and ground state
entropy of Ising spin models 
was first shown by Anderson.\cite{anderson}
Accurate calculations of thermodynamic properties of systems
with extensive ground state degeneracy\cite{liebmann,moessner,huber}
remains a challenging task. 

Interest in such systems has grown considerably with the discovery
of spin-ice materials.\cite{gingras} These are pyrochlore-lattice
spin systems with strong uniaxial anisotropy. In each tetrahedron of
the lattice, two spins point in and two point out,
providing an exact realization of Anderson's
spin-ice mapping. 
In the real materials, the spin-ice states arise
from a combination of nearest-neighbor exchange and long-range dipolar 
interactions.\cite{castelnovo,gingras} 
Due to the unusual angles between the easy-axis
directions at neighboring sites, antiferromagnetic exchange
leads to a lower energy for all-in/all-out configurations in a tetrahedron,
while ferromagnetic exchange leads to the two-in/two-out
spin-ice configurations. 
The measured residual entropy, in these systems, is in good 
agreement with the Pauling 
value.\cite{ramirez} 

In this paper, our focus is the study of an Ising antiferromagnet
on the pyrochlore lattice, which from a statistical mechanics point
of view is equivalent to a ferromagnetically
exchange coupled spin-ice material. 
It is defined by the Hamiltonian
\begin{equation}
{\cal H}=\sum_{i,j} S_i S_j,
\end{equation}
Here the spins $S_i$ take values $\pm 1$ and the sum runs over all
nearest-neighbor bonds of the pyrochlore lattice.
Single tetrahedron
based approximations for the thermodynamic properties of
this model are common.\cite{pauling,liebmann, moessner,berlinsky,castelnovo,castelnovo1} We note that single tetrahedron based approximation 
for residual entropy is not 
the same as the entropy of a single tetrahedron.
Translating Pauling's argument to the spin language, each
spin has two states, but 
in each tetrahedron only $6$ 
out of $16$ spin configurations obey the ice-rules.
Treating the constraint in each tetrahedron as independent,
a system with $N$ spins has
$2^{N}({6\over 16})^{N_t}$ ground states, where $N_t$ is number of tetrahedra.
Since, $N_t$ equals $N/2$, this leads to a ground state
entropy per spin of $S={1\over 2}\log{3\over 2}$.

Corrections to 
the Pauling expression for residual entropy  of ice
have been studied before.\cite{nagle}
Following earlier work by Takahashi\cite{takahashi} and
DiMargio and Stillinger,\cite{stillinger}
Nagle used graphical methods on Vertex
models\cite{nagle} to study ground state residual entropy for cubic
ice corresponding to the pyrochlore lattice as well as the
more common ice structure known as Layered Hexagonal ice.
He found that the difference in residual entropy between the
two structures was negilgible. The corrections to the Pauling
expression was estimated to be about $1.1$ percent.
Comparable corrections (approximately $1.2$ percent)
have been estimated in more recent
computational studies.\cite{berg}
Here, we will calculate the thermodynamic properties of
the antiferromagnetic Ising model on the pyrochlore lattice
using series expansion methods.\cite{book}

In a Numerical Linked Cluster expansion (NLC),\cite{rigol}
an extensive property $P$ for
a large lattice ${\cal L}$ of N-sites is expressed as
\begin{equation}
P({\cal L})/N=\sum_c L(c) \times W_P(c).
\end{equation}
The sum is over distinct clusters of the lattice. $L(c)$ is the lattice
constant of the cluster, given by the number of embeddings of the
cluster in the lattice, per site. The quantity $W_P(c)$ is the
weight of the cluster associated with the property $P$, which is
defined by the subgraph subtraction scheme
\begin{equation}
W_P(c)=P(c)-\sum_s W_P(s),
\end{equation}
where the sum runs over all subclusters of the cluster $c$. 
Thus, to carry out the calculation up to some order,
one needs a count of all the clusters up to some order, and
the property $P(c)$ needs to
be calculated for every cluster
to high precision using numerical methods.

The number of clusters, needed in the study, is significantly reduced
if it can be shown that only star-graphs contribute.\cite{mckenzie}
Consider a graph as a collection of sites that are connected
pair-wise by the bonds of the graph. A site is a point of articulation
in a graph, if cutting all the bonds incident on the site,
makes the rest of the graph disconnected.
A star-graph is one that has no such articulation site.
Star graph expansions have been used to develop high temperature
expansions for various classical spin models.\cite{mckenzie} But, they can also
be used in a Numerical Linked Cluster (NLC) scheme, where
no small expansion parameter is needed.\cite{rigol} Rather,
thermodynamic properties of finite clusters can be calculated
at a given temperature and then the principle of inclusion and
exclusion can be used to calculate the thermodynamic property
of the infinite system by summing up contributions
from all allowed clusters.

A star graph expansion requires that all articulated graphs
have zero weight. This is very simple to see for properties
that can be obtained from the logarithm of the zero-field
partition function. For any articulated graph, one can show
that a partition function, that is normalized to unity
in the absence of interactions, becomes a product of
partition functions over the two subgraphs articulated at a point.
Hence $\log{Z}$ becomes a sum. Thus for an articulated graph c
made up of parts a and b, which only share a point of articulation,
the property satisfies
\begin{equation}
P(c)=P(a)+P(b).
\end{equation}
This is enough to ensure that after subgraph
subtraction the articulated graph has zero weight.
It is a little harder to see how to
develop star-graph expansion for the uniform susceptibility.
One needs to consider the matrix $M$
with elements\cite{rapaport,singh-chakravarty}
\begin{equation}
M_{ij} =<S_i S_j>,
\end{equation}
where $i$ and $j$ are sites of the cluster and angular brackets
denote thermal expectation values.
One can show that for a graph articulated at site $k$ into parts $a$ and $b$
(with  spin normalization $ <S_k^2>=1$)
\begin{equation}
<S_i S_j>=<S_i S_k>_a\ <S_k S_j>_b,
\end{equation}
when $i$ and $j$ belong to $a$ and $b$ respectively. And
within a and b, the correlations remain the same as in
the subcluster. This can be used to further show 
that\cite{rapaport,singh-chakravarty}
\begin{equation}
\psi=\sum_{i,j} M_{ij}^{-1},
\end{equation}
where $M^{-1}$ is the inverse of the matrix $M$,
has a star graph expansion. Furthermore, for the infinite lattice,
this is related to the inverse of the uniform structure factor,
or temperature times the uniform susceptibility by the relation,
\begin{equation}
T\chi ={1\over N} \sum_{i,j} <S_i S_j>={N\over \psi}.
\end{equation}

\begin{figure}
\begin{center}
\includegraphics[width=6cm]{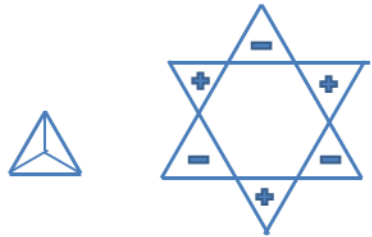}
\caption{\label{figa} (color online)
First two star-graphs of the pyrochlore lattice made up of
complete tetrahedra. The first one is a single tetrahedron. 
The second is a ring of six tetrahedra that alternately point
in and out of the board (denoted by $+$ and $-$ signs).
}
\end{center}
\end{figure}

In a lattice of corner-sharing
tetrahedra, it is natural to consider an NLC  expansion
scheme, in which all interactions are grouped into tetrahedra.
Thus, apart from a single site, all graphs are made up of 
complete tetrahedra. 
This scheme is particularly useful because the tetrahedra
are joined at corners in the lattice.
In a graph with two tetrahedra, the site where
they are joined becomes a point of articulation. Cutting all
the bonds at that site makes the graph disconnected. Thus,
in a star-graph expansion, the two-tetrahedron graph
makes no contribution. It also means that the single tetrahedron
approximation becomes exact on a Husimi tree of tetrahedra,\cite{husimi,oleg}
where there are no other closed loops of tetrahedra.

The tetrahedra of a pyrochlore lattice are known to form
a diamond lattice. Thus any graph counting problem involving
tetrahedra is equivalent to counting graphs on the
diamond lattice, where sites of the diamond lattice represent
tetrahedra while bonds of the diamond lattice represent shared sites
between neighboring tetrahedra. All star graphs on the diamond lattice up to
$16$ bonds have been listed by Sykes {\it et al}.\cite{sykes} We
will make use of these to calculate the expansions to $16$-th order.
One should note that all lattice constants of the diamond lattice
need to be divided by $2$, because the number of tetrahedra is
${1\over 2}$ the number of sites in the pyrochlore lattice.

We illustrate how the method works by showing the first two orders
of the calculation for the ground state entropy (with the
Boltzmann constant $k_B=1$). 
The first two star-graphs with complete tetrahedra are shown in Fig.~1.
They have a lattice constant per site of the pyrochlore lattice
of ${1\over 2}$ and $1$ respectively. One also needs to consider
a single-site, which provides all contributions before any 
interactions are included. It has a count of unity.

For our illustration, the ground state entropy is the property $P$.
In zeroth order, the single site has two ground states, giving an entropy
of $\log{2}$. It has no subgraphs. Hence, its weight is also $\log{2}$.
The first star graph, a single tetrahedron, has $6$ ground states.
Hence the property $P$ for the graph equals $\log{6}$.
To obtain its weight, one must subtract the weights of
the 4 sites. Thus the weight of the single tetrahedron is
$$W=\log{6}-4 \log{2}=\log{3/8}$$
Thus, to first order the ground state entropy, per site,
for the infinite system is
$$S=\log{2}+{1\over 2}\log{3/8}={1\over 2}\log{3/2}\approx 0.20273.$$
Note that the factor of ${1\over 2}$ in front of $\log{3/2}$ is the count of
number of tetrahedra per lattice site, which is one-half.
This is the Pauling answer. The next star graph is
a graph of $6$ tetrahedra (See Fig.~1). It has 730 ground states.
It has $6$ single tetrahedron subgraphs and 18 sites.
Thus, its weight is
$$W=\log{730}-6\log{3/8}-18\log{2}=\log{730/729}$$
Thus, to this order, the entropy per site, becomes
$$S=\log{2}+{1\over 2}\log{3/8}+\log{730/729}\approx 0.20410.$$

These first corrections are analogous to triangle-based
NLC calculations done by Rigol et al for the kagome 
lattice.\cite{rigol} In that case, the Pauling expression
for ground state entropy is $0.50136$. Adding the next
correction brought the entropy to $0.50182$, much closer to
known exact answer of approximately $0.50183$.\cite{exact-kagome}
Because the count of longer loops increases much more rapidly
in a 3-dimensional lattice, a high order calculation is
needed to assess the corrections more accurately.
Here we have done only the first corrections for the
uniform susceptibility, but a $16$-th order correction
for the entropy and specific heat.

\begin{figure}
\begin{center}
\includegraphics[width=7cm,angle=270]{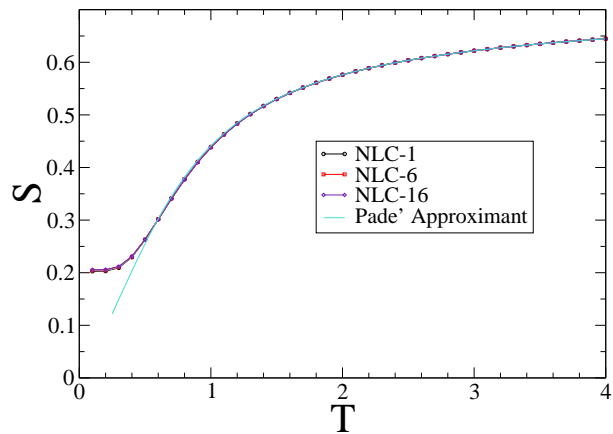}
\caption{\label{figa} (color online)
Entropy of the pyrochlore lattice Ising antiferromagnet
as a function of temperature.
}
\end{center}
\end{figure}

\begin{figure}
\begin{center}
\includegraphics[width=7cm,angle=270]{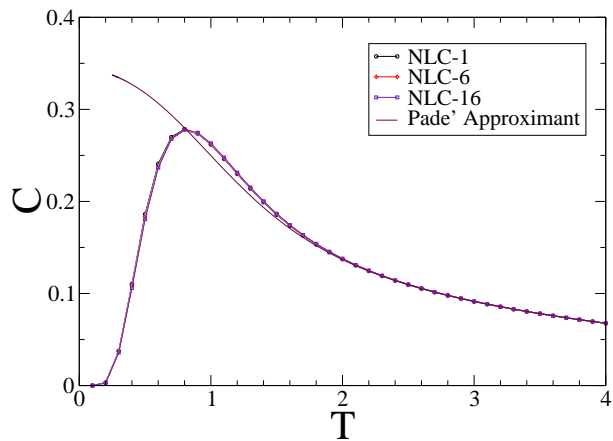}
\caption{\label{figb} (color online)
Heat capacity of the pyrochlore lattice Ising antiferromagnet
as a function of temperature.
}
\end{center}
\end{figure}

High temperature series expansions for this system were derived
some years ago: to order $19$ for $\log{Z}$ (See Ref.~\cite{gibberd}) and
to order $16$ for $\chi$.\cite{oitmaa} We have reanalyzed these
series using Pad\'e and d-log Pad\'e approximants to yield the
series results for comparison at high temperatures.

Fig.~2 shows a plot of the entropy. The single tetrahedra
approximation is denoted NLC-1. NLC-6 includes
the next order correction. NLC-16 gives the result up to 16th order.
Fig.~3 shows the corresponding
plots of the heat capacity. Fig.~4 shows plots of the
uniform susceptibility. 
The high temperature expansions converge really well
only above a temperature of $2$.
Note that the spins are normalized to be $\pm 1$. The temperature
scale would be four times lower if they were normalized to $\pm 1/2$.

In all cases, the single tetrahedron
based approximation is quite accurate. 
Let us define percent
correction for the three quantities $a=$ entropy ($s$), heat capacity ($c$)
and susceptibility ($\chi$) as
\begin{equation}
P_a = 100*(a(N)-a(1))/a(1).
\end{equation}
Here $a(N)$ is the quantity after $N$th order NLC.
and $a(1)$ is the property with only the one tetrahedron cluster.
A plot of $P_a$ with temperature is shown in Fig.~5. The largest
corrections are near $T=0$. They are a little over one percent for
the entropy and about $5$ percent for the specific heat.

At $T=0$, our graphical scheme, reduces to one of counting ground states
on increasingly larger clusters. While this scheme is not identical to
the one used by Nagle,\cite{nagle} it is closely related. As found by Nagle,
the first few orders give the same answer for the cubic and 
layered hexagonal structures of ice. We also find that the first corrections
to the Pauling entropy are identical for the two lattices.
Since, our interest is in the
pyrochlore lattice spin model, we have not studied the
layered hexagonal  structure in higher orders.

\begin{figure}
\begin{center}
\includegraphics[width=7cm,angle=270]{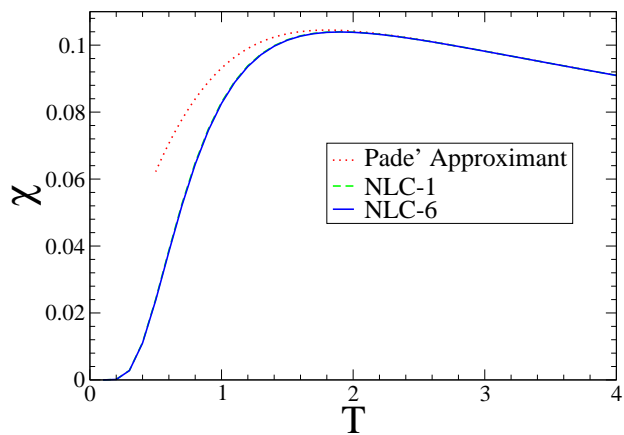}
\caption{\label{figc} (color online)
Uniform susceptibility of the pyrochlore lattice Ising antiferromagnet
as a function of temperature. Note that NLC-1 and NLC-6 can barely
be differentiated in the plot.
}
\end{center}
\end{figure}

\begin{figure}
\begin{center}
\includegraphics[width=7cm,angle=270]{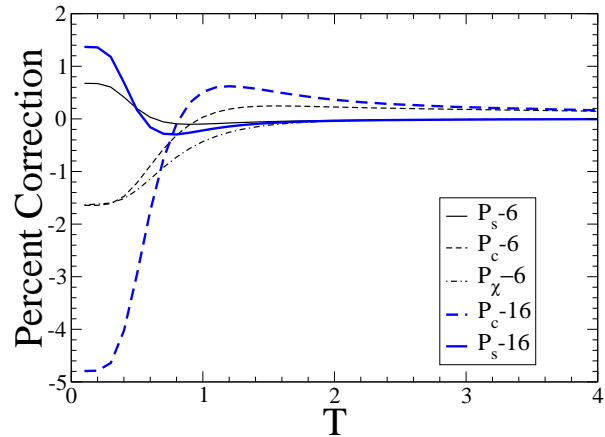}
\caption{\label{figd} (color online)
Percentage correction to the quantities, entropy ($P_s$),
heat capacity ($P_c$) and susceptibility ($P_\chi$) after
$6$th and $16$th order NLC calculations (denoted by the
numbers $6$ and $16$ in the legends).
}
\end{center}
\end{figure}

In conclusion, we have used a star-graph expansion scheme to
show that the Pauling approximation for the entropy of ice and
calculation of other properties of the pyrochlore lattice Ising
model based on a single tetrahedron
is really a first term in a Numerical Linked Cluster (NLC)
scheme. In the corner sharing lattice, this scheme 
is highly accurate. We have calculated corrections
to the single tetrahedron approximations and found them
to be at most a few percent for different thermodynamic quantities.

These ideas of star-graph expansions are also valid
for continuous spin models, where also single tetrahedra
based schemes have been used successfully.\cite{berlinsky} 
However, it should be noted that finite-size calculations
of properties of the continuous spin systems can be a non-trivial task.
Star graph expansions are also valid in presence of dilution and
quenched disorder.\cite{ditzian,singh-chakravarty}
Star graph
expansions are not valid for quantum spin models and
hence all connected clusters of tetrahedra need to be
included in these cases.

\begin{acknowledgements}
We are grateful for the computing resources provided by the 
Australian Partnership for Advanced Computing (APAC) National Facility.
This work is supported in part by NSF grant number  DMR-1004231.
\end{acknowledgements}


\begin{thebibliography}{2}

\bibitem{pauling} L. Pauling, J. Am. Chem. Soc. 57, 2680 (1935).

\bibitem{wannier} G. H. Wannier, Phys. Rev. 79, 357 (1950).

\bibitem{anderson} P. W. Anderson, Phys. Rev. 102, 1008 (1956).

\bibitem{liebmann}
R. Liebmann, {\it Statistical Mechanics of
Periodic Frustrated Ising Systems},
(Berlin: Springer-Verlag 1986)

\bibitem{moessner} R. Moessner, Can. J. Phys. 79, 1283 (2001).

\bibitem{huber} A.J. Garcia-Adeva and
D.L. Huber Phys. Rev. Lett. 85, 4598 (2000).

\bibitem{gingras} For a recent review of
spin-ice see J. S. Gardner, M. J. P. Gingras and J. E. Greedon,
Rev. Mod. Phys. 82, 53 (2010).

\bibitem{castelnovo} C. Castelnovo, R. Moessner, and S. L. Sondhi,
Nature 451, 42 (2008).

\bibitem{ramirez} A. P. Ramirez et al, Nature 399, 333 (1999).

\bibitem{berlinsky} R. Moessner and A. J. Berlinsky, Phys. Rev. Lett.
83, 3293 (1999).

\bibitem{castelnovo1} C. Castelnovo, R. Moessner, and S. L. Sondhi,
Phys. Rev. B 84, 144435 (2011).

\bibitem{nagle} J. F. Nagle, J. Math Phys. 7, 1484 (1966).

\bibitem{takahashi} H. Takahashi, Proc. Phys. Math. Soc. (Japan)
23, 1069 (1941).

\bibitem{stillinger} E. A. DiMarzio and F. H. Stillinger, Jr.,
J. Chem. Phys. 40, 1577 (1964).

\bibitem{berg} B. A. Berg, C. Muguruma and Y. Okamoto, Phys. Rev.
B 75, 092202 (2007).

\bibitem{book} J. Oitmaa, C. Hamer and W. Zheng, {\it Series Expansion
Methods for strongly interacting lattice models} (Cambridge University
Press, 2006).

\bibitem{rigol} M. Rigol, T. Bryant and R. R. P. Singh,
Phys. Rev. Lett. 97, 187202 (2006); Phys. Rev. E 75, 061118 (2007);
Phys. Rev. E 75, 061119 (2007).

\bibitem{mckenzie} For a review of star-graph expansions, see
S. Mckenzie in {\it Phase Transitions Cargese} (1980), edited by M. Levy,
J. C. LeGuillou, and J. Zinn-Justin (Plenum, New York, 1980).

\bibitem{rapaport} D. C. Rapaport, J. Phys. A 7, 1918 (1974).

\bibitem{singh-chakravarty} R. R. P. Singh and S. Chakravarty,
Phys. Rev. Lett. 57, 245 (1986);
Phys. Rev. B 36, 546 (1987).


\bibitem{husimi} K. Husimi, J. Chem. Phys. 18, 682 (1950).

\bibitem{oleg} Z. Hao and O. Tchernyshyov, Phys. Rev. Lett. 81,
214445 (2010).

\bibitem{sykes} M. F. Sykes, J. W. Essam, B. R. Heap and
B. J. Hiley, J. Math. Phys. 7, 1557 (1966).


\bibitem{exact-kagome}
I. Syozi, Prog. Theor. Phys. {\bf 6}, 306 (1951);
K. Kano and S. Naya, ibid {\bf 10}, 158 (1953).

\bibitem{gibberd} R. W. Gibberd, Can J. Phys. 48, 307 (1970).

\bibitem{oitmaa} J. Ho-Ting-Hun and J. Oitmaa, J. Phys. A 8, 1920 (1975).

\bibitem{ditzian} R. V. Ditzian and L. P. Kadanoff, Phys. Rev. B 19,
4631 (1979).

\end{thebibliography}

\end{document}